# Direct detection of polar structure formation in helium nanodroplets by beam deflection measurements

John W. Niman,[a] Benjamin S. Kamerin,[a] Lorenz Kranabetter,[b] Daniel J. Merthe,[a],†
Jiří Suchan,[c] Petr Slavíček,*[c,d] Vitaly V. Kresin*[a]

[a] *Department of Physics and Astronomy, University of Southern California, Los Angeles, CA 90089-0484, USA*

[b] *Institut für Ionenphysik und Angewandte Physik, Universität Innsbruck, Technikerstr. 25, A-6020 Innsbruck, Austria*

[c] *Department of Physical Chemistry, University of Chemistry and Technology, Technická 5, Prague 6, Czech Republic*

[d] *J. Heyrovský Institute of Physical Chemistry v.v.i., The Czech Academy of Sciences, Dolejškova 3, 18223 Prague, Czech Republic*

**Abstract**

Long-range intermolecular forces are able to steer polar molecules submerged in superfluid helium nanodroplets into highly polar metastable configurations. We demonstrate that the presence of such special structures can be identified, in a direct and determinative way, by electrostatic deflection of the doped nanodroplet beam. The measurement also establishes the structures' electric dipole moments. In consequence, the introduced approach is complementary to spectroscopic studies of low-temperature molecular assembly reactions. It is enabled by the fact that within the cold superfluid matrix the molecular dipoles become nearly completely oriented by the applied electric field. As a result, the massive (tens of thousands of helium atoms) nanodroplets undergo significant deflections. The method is illustrated here by an application to dimers and trimers of dimethyl sulfoxide (DMSO) molecules. We interpret the experimental results with ab initio theory, mapping the potential energy surface of DMSO complexes and simulating their low temperature aggregation dynamics.

**Introduction**

Long-range intermolecular forces play an essential role in reactions at sub-Kelvin temperatures (see, e.g., the reviews in refs [1-4). For example, long-range interactions between polar molecules embedded in helium nanodroplets often dominate the outcome of their assembly reactions. This is facilitated by the low internal temperature (370 mK) of the nanodroplet medium as well as by its superfluidity.[5] As a result, molecular reorientation and intermolecular reactions within nanodroplets are not perturbed by inhomogeneities present in other low-temperature surface and matrix isolation environments, making these "nano-cryo-traps" excellent hosts for exploring the physics and chemistry of cold molecular systems.[6]

A landmark demonstration of the action of long-range forces was furnished by experiments on HCN molecules sequentially picked up by a He nanodroplet beam.[7] These linear molecules were guided by dipole-dipole forces to self-assemble into long chains aligned head-to-tail inside the nanodroplet. HCCCN was found to behave similarly.[8] These chains rank among the most polar molecular systems ever observed in a molecular beam. In an "ordinary" environment thermal motion would drive them out of this type of metastable configuration, but within a very cold and inert liquid helium droplet they become long-lived. Data on formic acid,[9] imidazole,[10] and acetic acid[11,12] dimers suggested an analogous alignment mechanism.

However, such an outcome is not universal in nanodroplet embedding. For example, two HCl molecules arrange themselves nearly at a right angle to each other[13,14] while water clusters form cyclic structures.[15] The "decision" by polar molecules how to orient themselves upon approach depends on the strength of their dipoles, on their responsiveness to the mutually reorienting torques (i.e., their rotational constants and their accessible rotational quantum states), and on the directionality and flexibility of their bond formation. That is to say, the outcome depends on the shape of the intermolecular potential energy surface and on the barrier heights encountered on the path to the final configuration.

It is therefore interesting and informative to establish whether a molecular formation within a nanodroplet can reach its global energy minimum or finds itself trapped in a polar metastable state. However, often this is not a straightforward determination. The studies cited above based their conclusions on the interpretation of dopant infrared spectra or on inference from electron attachment mass spectrometry. Such assignments grow more difficult and less definitive with increasing size and/or complexity of the embedded molecules and their assemblies.

In this work we describe a measurement which *directly* establishes the polarity of a molecular assembly, as well as determines its dipole moment. It makes use of electrostatic deflection of the doped nanodroplet beam.[16,17]

The technique is based on the fact that polar structures embedded within the superfluid matrix can be made nearly fully oriented by an external static electric field[18] and consequently experience an extremely large deflecting force from the field's gradient. Such a high degree of orientation is unattainable for bare polyatomic complexes in a molecular beam. Whereas some relatively small and light molecules reach rotational temperatures $T_{rot}$ below 1 K with the use of seeded supersonic expansions and exhibit large deflections (see, e.g., refs [19,20]), this becomes impractical for heavier systems.

For the purpose of an estimate, consider the classical Langevin function for the orientation of a molecular rotor in an external field $E\hat{z}$: $\bar{p}_z/p_0 = [\coth x - 1/x]$. This is a good approximation[21,22] for $k_B T_{rot} \gg B$. Here $p_0$ is the molecule's dipole moment, $\bar{p}_z$ is the average projection of this dipole on the field axis, $x \equiv p_0 E / k_B T_{rot}$, and $B$ is the rotational constant. For $T_{rot}$ above a few K and practical electric field strengths, the ratio $x$ remains small even for dipole moments of several Debye (D), and in this limit $\bar{p}_z/p_0 \approx x/3 \ll 1$. Therefore it is only when the rotational temperature becomes very low, as enabled in the present case by helium nanodroplet isolation, that the orientation can approach saturation ($\bar{p}_z \to p_0$). This effect has been taken advantage of in landmark experiments using pendular-state spectroscopy.[18]

If the external electric field which orients the nanodroplet-submerged dipoles is designed also to have a strong gradient in the same direction, then these dipoles will experience such a strong sideways force $F_z = p_z \left(\partial E/\partial z\right)$ that the massive doped droplets, comprised of tens of thousands of helium atoms, will be significantly deflected in their entirety. Thus, our procedure involves comparing the deflection profile of a singly-doped nanodroplet beam with that of a beam composed of multiply-doped nanodroplets. If, for example, the droplets containing two (or three, etc.) molecules show negligible deflection, we can immediately conclude that the dimer (trimer, etc.) has settled into a nonpolar configuration. A strongly deflected profile, on the other hand, immediately attests to the formation of a polar structure, and the magnitude of the deflection translates into the magnitude of this formation's total dipole moment.

This is a conveniently unambiguous measurement applicable to a wide range of molecules, from diatomic to polyatomic (including biological). Practically any molecular species that can be brought into the vapor phase with a pressure of only $10^{-6}$-$10^{-4}$ mbar can be picked up by the nanodroplet beam and thermalized within the inert viscosity-free medium. The thermalization proceeds by evaporative cooling: the molecules' translational and internal energies are transferred to the superfluid matrix which has a very high thermal conductivity, and released via evaporation of surface helium atoms, promptly bringing the nanodroplet back to the original temperature.[5]

Here we apply the deflection method to monomers, dimers and trimers of the dimethyl sulfoxide molecule ("DMSO," $(CH_3)_2SO$, molecular mass 78 Da). The molecule is nearly an oblate symmetric top, with rotational constants of[23,24] 0.235 cm$^{-1}$, 0.231 cm$^{-1}$, and 0.141 cm$^{-1}$ and its total dipole moment is[25] *p*=4.0 D. The measurement clearly reveals the presence of highly polar dimers and trimers, i.e., the formation of metastable polar configurations abetted by the cryogenic nanodroplet environment. To our knowledge, this is the first direct non-spectroscopic identification of such a cold polar molecular assembly.

**Results and Discussion**

*Deflection profiles.* The experimental setup has been described in detail elsewhere.[16,17,26] A nanodroplet beam is formed by cold nozzle expansion of pure helium gas. It passes first through a pick-up cell filled with DMSO vapor, and then between two high-voltage electrodes which create an electric field and a collinear field gradient directed perpendicular to the beam axis. Downstream, the beam enters through a slit into an electron-impact ionizer, and the intensities of the resulting molecular ions are recorded by a quadrupole mass spectrometer in synchronization with a beam chopper. The deflection induced by the electric field is determined by comparing the beam's "field-on" and "field-off" spatial profiles which are mapped out by translating the detector chamber, with its entrance slit, on a precision linear stage.

Molecules are picked up by helium nanodroplets via successive collisions in a Poisson process.[5] Therefore it is important to correlate measured beam deflections with the specific number of molecules embedded in the droplet. In other words, when mapping out the deflection profile of a dopant ion peak in the mass spectrum, we need to ensure that it is not a fragment of a larger agglomerate. This is done by gradually increasing the vapor pressure in the pick-up cell and monitoring the mass spectrum for the appearance of molecular ions characteristic of progressively larger entities. For example, monomer ionization produces a strong (DMSO)$^+$ signal[27] at *m*=78 Da, hence if we measure beam profiles with the mass spectrometer set to this mass peak but with the vapor pressure low enough to suppress the corresponding characteristic (DMSO)$_2^+$ peak at $m$=156 Da, then we can be confident that the deflection principally corresponds to droplets carrying the monomer. Similarly, profiles measured at $m$=156 Da but before the appearance of the trimer's signal must derive from the dimer, etc. Representative mass spectra are shown in the Supporting Information (SI).

Fig. 1 shows the deflection profiles of helium nanodroplets containing one, two, and three DMSO molecules. The deflections are substantial despite the fact that the droplets are truly massive (~$1\times10^4$–$3\times10^4$ He atoms, as described in the caption). Therefore we are immediately and directly informed by Fig. 1(b) that (DMSO)$_2$ settles into a strongly polar configuration and not into its global minimum structure, because the latter would be symmetric with a zero dipole moment.[28]

In order to assign an absolute value of the dipole moment to the dopant, we must keep in mind that the host nanodroplets are not all of the same size. The size distribution produced by the nozzle expansion is log-normal, and this translates into a convolution of pick-up cross sections, deflection angles, and ionization efficiencies. Our procedure[16,17] is to start with the profile corresponding to a single DMSO dopant molecule whose dipole moment is known. A fit to the deflected profile (by a Monte Carlo simulation of the pick-up, evaporation, deflection, and detection steps) is used to calibrate the droplet size distribution. Then by repeating the deflection measurement and its simulation with doubly- and triply-doped nanodroplets produced and detected under the same conditions, we can deduce the electric dipole moments corresponding to the dimer and the trimer.

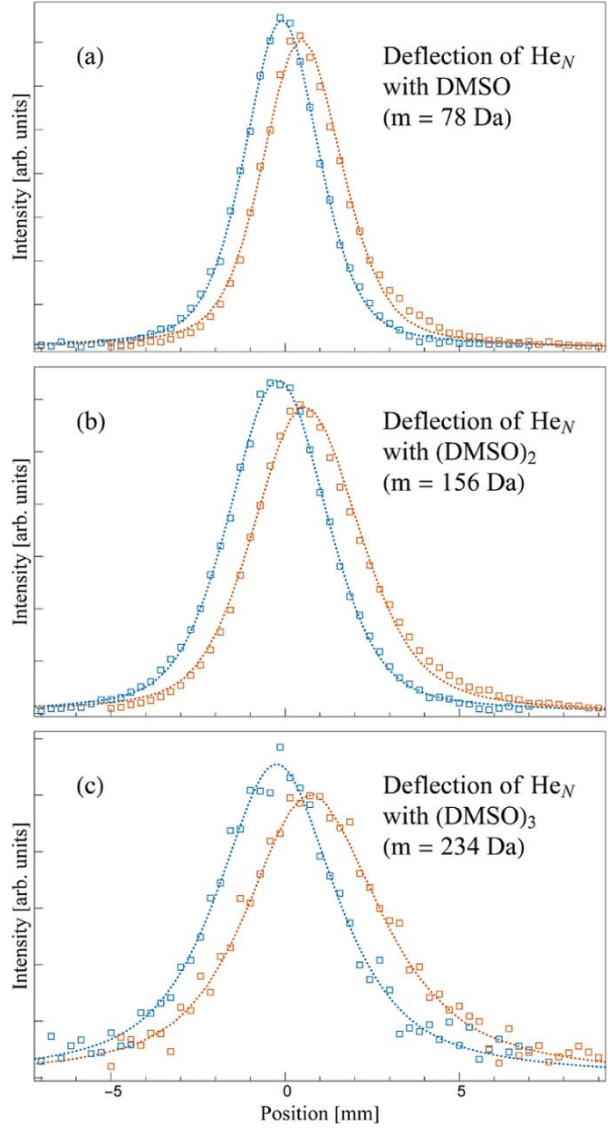

**Figure 1.** Profiles of (DMSO)$_n$-doped helium nanodroplet beams. Blue: zero-field profiles, orange: deflection by a field of 82 kV/cm strength and 338 kV/cm$^2$ gradient. Symbols: experimental data, lines: fits of the deflection process, as described in the text. The monomer profile mapped for a particular temperature $T$ and stagnation pressure $P$ of the He$_N$ beam source is used to determine the average $\bar{N}$ and width $\Delta N$ of the nanodroplet size distribution, and then fits to the dimer and trimer profiles for the same source conditions yield these dopants' dipole moments. In (b) $P$=80 bar, $T$=15.5 K, $\bar{N} \approx 2.3 \times 10^4$, in (c) $P$=80 bar, $T$=16.4 K, $\bar{N} \approx 1.4 \times 10^4$.

These dipole moments enter the fitting procedure at the step where the deflecting electrostatic force is calculated. As described in the Introduction, this requires knowing $\bar{p}_z$, i.e., the degree of orientation induced by the applied field. For the DMSO monomer this is carried out by diagonalizing the rotational Stark effect matrix (cf. ref [29]) using the components of the molecule's dipole moment.[24] For the heavier dimer and trimer the classical Langevin-Debye formula is sufficiently accurate.[30] In calculating the monomer's Stark spectra one should keep in mind that rotational coupling to the superfluid[31] enhances the moments of inertia of the heavier molecular rotors by an average factor of ~2.5-3 compared with their gas phase value.[5,18] Since DMSO's specific renormalization factor is not known, it was set to 2.6 in our data fitting procedure. We found that the inclusion of this factor had practically no effect on the deduced dipole of the dimer but shifted that of the trimer upward by ≈10%-15%. For the final fitted dipole values listed below, the (DMSO)$_n$ orientations within an applied 82 kV/cm field were found to be 86%, 97%, and 98% for $n$=1-3, respectively.

*Dipole moments.* From analysis of the measurements, we assign effective electric dipole moments of 7.2 D to (DMSO)$_2$ and 8.6 D to (DMSO)$_3$, with an estimated accuracy of ±0.2 D and ±0.6 D, respectively. These values, which can be compared with the ground state moments of 0 D for the aforementioned symmetric dimer and 4.2 D for the trimer[28] (essentially a nonpolar dimer plus an unpaired monomer), establish the presence of highly polar metastable structures. In the cold superfluid environment these structures are steered into formation by the long-range intermolecular forces and are then unable to overcome the potential barrier leading to the global minimum configuration.

*Modeling of molecular complex formation.* To facilitate the interpretation of the above results, we supplemented the experiments with *ab initio* modeling of DMSO condensation. We optimized the geometry of DMSO dimers and trimers with the B3LYP functional with the aug-cc-pVDZ basis set. The DMSO complexes are dominantly bound by electrostatic forces but the dispersion interactions still play a non-negligible role. We have therefore used the D2 correction of Grimme.[32] The approach was tested against the CCSD(T)/aug-cc-pVTZ method for the DMSO dimer, yielding similar energetics (see the SI). All calculations were performed in the gas phase: by considering complexes with helium atoms or within a dielectric continuum we found that the helium environment had a negligible effect on the structure and energetics. The potential energy surfaces (PES) were pre-screened with molecular mechanics (MM)-based metadynamics simulations[33] and the structures were then recalculated at the DFT level (see the SI for further information).

The process of DMSO dimer formation was modeled with molecular dynamics (MD) simulations within the canonical ensemble. We used the Nosé-Hoover thermostat with a rather small value of $\tau = 0.01$ ps. This corresponds to fast draining of extra energy from the system, so that at each time it essentially remains in equilibrium. A temperature of 5 K was chosen in order to accelerate the simulations. It is higher than in the experiment but the difference is small compared with the PES accuracy.

We started with two DMSO molecules positioned at a distance of 20 Å between the two sulphur atoms with a random orientation. We then performed molecular mechanics simulations with the MM force field.[34] The molecules gradually approached each other while aligning their dipole moment. Since the MM force field does not reproduce the energetics of the minima sufficiently well, at the intermolecular distance of 10 Å we reset the simulations, switching from the force field to the more accurate semiempirical density functional tight binding (DFTB) method[35] with D3 dispersion correction.[36,37] The system then continued to evolve in time for another 500 ps with a time step of 1 fs, using the velocity Verlet integrator. Dipoles along the path were recalculated at the B3LYP/aug-cc-pVDZ level.

The DFT and CCSD(T) calculations were performed in Gaussian09.[38] Molecular dynamics simulations were performed in GROMACS 2018.4[39] and the DFTB simulations in the DFTB+ 18.2 code.[35] We also utilized our in-house MD code ABIN.[40]

*Results of modeling.* Fig. 2 shows several low-lying minima of the DMSO dimer obtained from extensive mapping of its potential energy surface. The structures are divided into two classes of minima: non-polar and polar. The global minimum (complex D1) of (DMSO)$_2$ has a symmetrical configuration with a zero dipole moment, consistent with the aforementioned work.[28] Structures D2 and D3 also belong to the low dipole manifold. Complexes D4 and D5 represent polar type structures. The experimental data suggest that the highly polar structure D5, with an almost orthogonal arrangement of dipoles, predominantly forms within nanodroplets. It is separated from the global minimum by a barrier of 0.08 eV (see the SI), which is more than sufficient to prevent a D5 → D1 transition.

Structure formation under cryogenic conditions is therefore likely to proceed as follows. At large separation the dominant force is the dipole-dipole interaction which aligns the two DMSO molecules. As described in the SI, there is a barrierless pathway between this structure and the D5 minimum. Therefore the molecules approach each other gradually within the helium environment to which all excess energy is almost immediately drained. The (DMSO)$_2$ ends up trapped within the basin of complex D5.

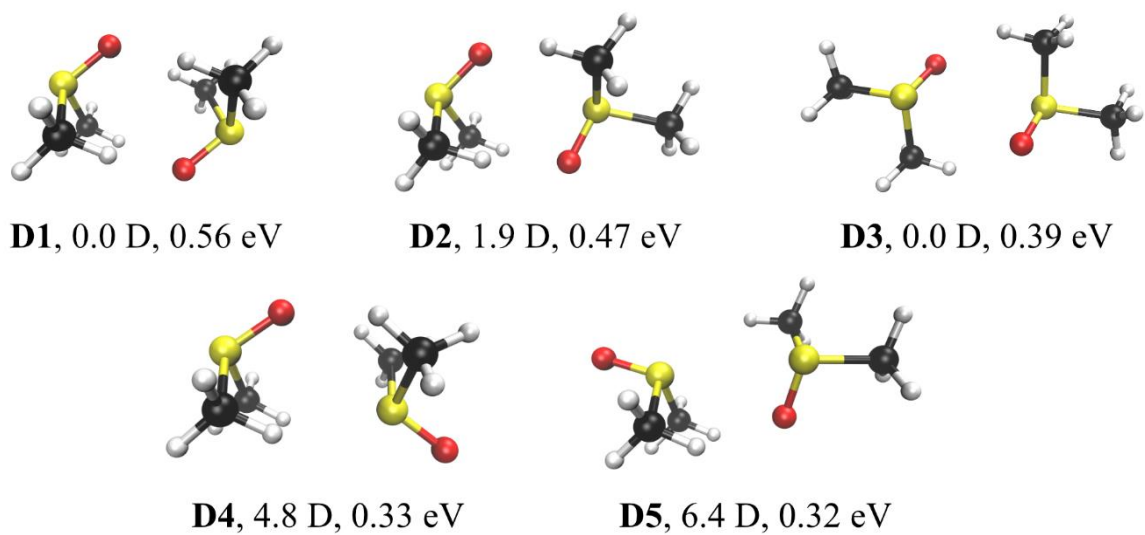

**Figure 2.** Energy minima of the DMSO dimer, with their corresponding binding energies and dipole moments.

We support this scenario by molecular dynamics (MD) simulations of the binary encounter under conditions of very efficient energy transfer. At the start the two dipoles are assigned a random relative orientation, but the trajectory shown in Fig. 3 demonstrates that it becomes correlated already at large distances. At closer approach the total dipole moment transiently increases. The molecular dipoles at that point are still parallel, hence the bump in the dipole moment is caused by mutual induction. Finally, the dimer quenches into one of the potential minima. In accord with the experiment, no formation of a zero dipole structure is found. The majority of the trajectories end up in the D5 minimum with a dipole of 6.4 D, some of them end up in the D4 minimum with a somewhat lower dipole moment than detected in the experiment.

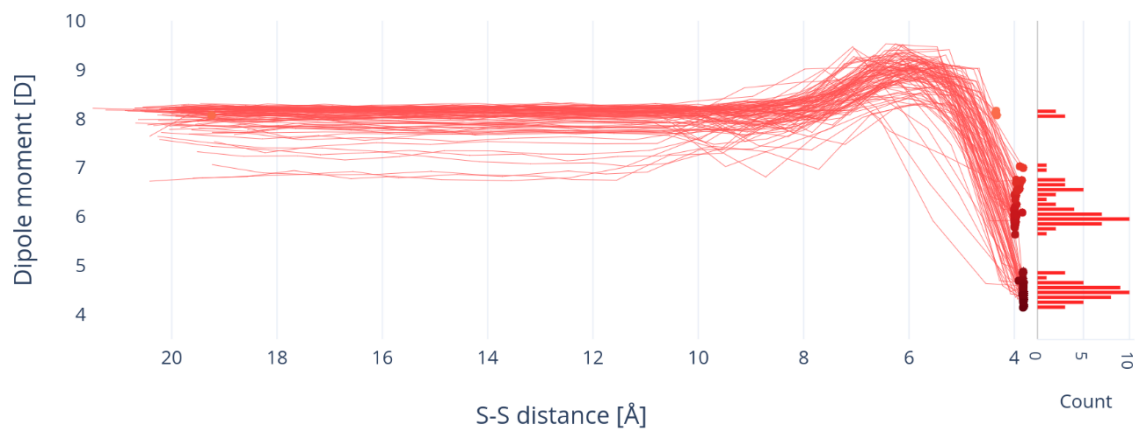

**Figure 3**. Dipole moment of DMSO dimer complex along the intermolecular approach coordinate, as illustrated by a molecular dynamics simulation.

The structures are more diverse for the trimer (Fig. 4). The lowest energy structure is cyclic with a dipole moment of 4.25 D (complex T1). Its formation is kinetically hindered. Indeed, as mentioned above, it represents the global dimer minimum to which the third molecule is added; since in the nanodroplets the former structure is not formed, neither will the cyclic trimer. We have located linear structures (T6, T7) with a much higher dipole close to 10 D. There are multiple other minima with intermediate dipoles. It follows from our simulations that a rather complex mixture of these metastable structures may be formed under the experimental conditions, and its precise assignment is beyond the reach of theory. The effective dipole moment of ≈8.6 D deduced from the deflection experiment represents the population average of the kinetically accessible structures.

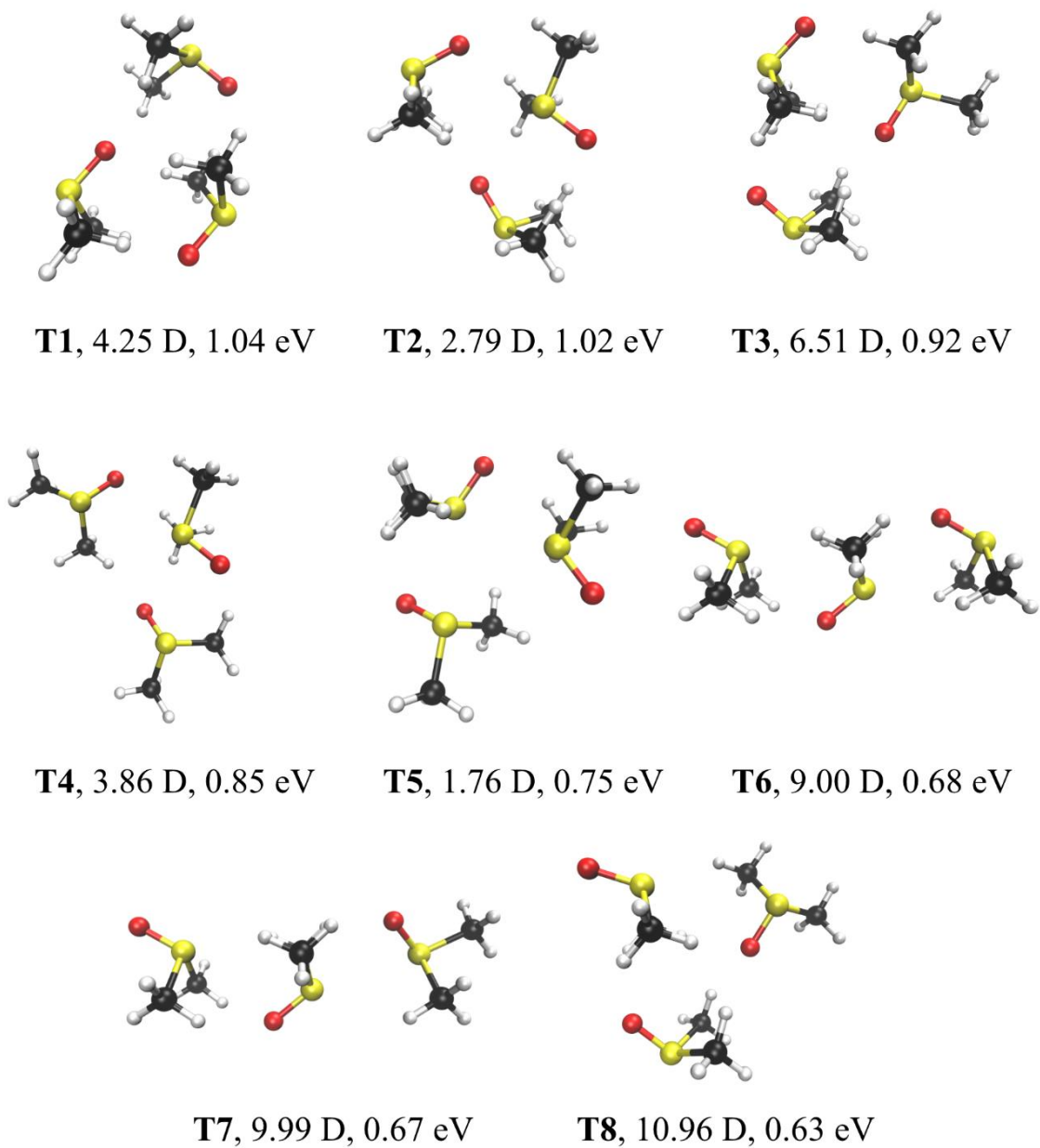

**Figure 4.** Energy minima of DMSO trimers, with their corresponding binding energies and dipole moments.

**Conclusion**

In summary, we have demonstrated that the presence of peculiar polar structures, formed by sequential embedding of polar molecules into superfluid helium nanodroplets, can be clearly and directly detected by electrostatic deflection of the doped nanodroplet beam. In an application of this method to DMSO molecules we found that they form dipole-aligned dimer and trimer structures, steered by long-range electrostatic interactions. The formation mechanism and the magnitudes of the dipole moments are in good agreement with calculations describing molecular interactions and structure formation in the viscosity-free cryogenic environment.

In future applications it will be interesting to extend this approach, for example, to a study of interactions between polar amino acids or between prototype solute and solvent molecules, as well as between molecules in photoinduced polar conformations. It is also interesting to inquire whether transfer of angular momentum between the impurities and the quantum-fluid bath, a phenomenon predicted to have the potential to screen the impurity – electric field interaction,[41] may be able to measurably affect the dynamics of molecular assembly within nanodroplets.

**Acknowledgments**

This work was supported by the U. S. National Science Foundation under Grant No. CHE-1664601. L.K. acknowledges a scholarship from the Austrian Marshall Plan Foundation and support from the Austrian Science Fund under project FWF W1259. J.S. and P.S. thank the Czech Science Foundation for support under Project number 18-16577S. J.S. is an International Max Planck Research School for Many Particle Systems in Structured Environments student. We would like to thank Jiahao Liang and Atef Sheekhoon for assistance.

**References**

[1]  M. T. Bell, T. P. Softley, *Mol. Phys.* **2009**, *107*, 99-132.

[2]  M. Schnell, G. Meijer, *Angew. Chem. Int. Ed.* **2009**, *48*, 6010-6031.

[3]  G. Quéméner, P. S. Julienne, *Chem. Rev.* **2012**, *112*, 4949-5011.

[4]  N. Balakrishnan, *J. Chem. Phys.* **2016**, *145*, 150901.

[5]  J. P. Toennies, A. F. Vilesov, *Angew. Chem. Int. Ed.* **2004**, *43*, 2622-2648.

[6]  K. K. Lehmann, G. Scoles, *Science* **1998**, *279*, 2065-2066.

[7]  K. Nauta, R. E. Miller, *Science.* **1999**, *283*, 1895-1897.

[8]  K. Nauta, D. T. Moore, R. E. Miller, *Faraday Discuss.* **1999**, *113*, 261-278.

[9]  F. Madeja, M. Havenith, K. Nauta, R. E. Miller, J. Chocholoušová, P. J. Hobza, *Chem. Phys.* **2004**, *120*, 10554-10560.

[10] M. Y. Choi, R. E. Miller, *J. Phys. Chem. A.* **2006**, *110*, 9344-9351.

[11] F. Ferreira da Silva, S. Jaksch, G. Martins, H. M. Dang, M. Dampc, S. Denifl, T. D. Märk, P. Limão-Vieira, J. Liu, S. Yang, A. M. Ellis, P. Scheier, *Phys. Chem. Chem. Phys.* **2009**, *11*, 11631-11637.

[12] J. A. Davies, M. W. D. Hanson-Heine, N. A. Besley, A. Shirley, J. Trowers, S. Yang, A. M. Ellis, *Phys. Chem. Chem. Phys.* **2019**, *21*, 13950-13958.

[13] M. Ortlieb, Ö. Birer, M. Letzner, G. W. Schwaab, M. Havenith, *J. Phys. Chem. A.* **2007**, *111*, 12192-12199.

[14] D. Skvortsov, R. Sliter, M. Y. Choi, A. F. Vilesov, *J. Chem. Phys.* **2007**, *128*, 094308.

[15] K. Nauta, R. E. Miller, *Science.* **2000**, *287*, 293-295.

[16] D. J. Merthe, V. V. Kresin, *J. Phys. Chem. Lett*. **2016**, *7*, 4879-4883.

[17] J. W. Niman, B. S. Kamerin, D. J. Merthe, L. Kranabetter, V. V. Kresin,. *Phys. Rev. Lett*., in press (July 2019).

[18] M. Y. Choi, G. E. Douberly, T. M. Falconer, Lewis, W. K. Lewis, C. M. Lindsay, J. M. Merritt, P. L. Stiles, R. E. Miller, *Int. Rev. Phys. Chem.* **2006**, *25*, 15-75.

[19] Y.-P. Chang, D. Horke, S. Trippel, J. Küpper, *Int. Rev. Phys. Chem*. **2015**, *34*, 557-590.

[20] M. Johny, J. Onvlee, T. Kierspel, H. Bieker, S. Trippel, J. Küpper, *Chem. Phys. Lett.* **2019**, *721*, 149-152.

[21] B. Friedrich, D. Herschbach, *Int. Rev. Phys. Chem.* **1996**, *15*, 325-344.

[22] J. Bulthuis, J. A. Becker, R. Moro, V. V. Kresin, *J. Chem. Phys.* **2008**, *129*, 024101.

[23] W. Feder, H. Dreizler, H. D. Rudolph, V. Typke, *Z. Naturforsch.* **1969**, *24a*, 266-278.

[24] M. L. Senent, S. Dalbouha, A. Cuisset, D. Sadovskii, *J. Phys. Chem. A.* **2015**, *119*, 9644-9652.

[25] *CRC* Handbook *of Chemistry and Physics*, 99[th] ed. (Ed.: J. R. Rumble), CRC Press, Boca Raton, **2018.**

[26] D. J. Merthe, PhD dissertation, University of Southern California (Los Angeles), **2017**.

[27] *NIST Chemistry WebBook*, *NIST Standard Reference Database Number 69* (Eds.:P. J. Linstrom, W. G. Mallard), National Institute of Standards and Technology, Gaithersburg, **2018**. http://webbook.nist.gov.

[28] N. S. Venkataramanan, A. Suvitha, Y. Kawazoe, *J. Mol. Liq.* **2018,** *249*, 454-462.

[29] Y.-P. Chang, F. Filsinger, B. G. Sartakov, J. Küpper, *Comput. Phys. Commun.* **2014**, *185*, 339-349.

[30] L. Pei, J. Zhang, W. Kong, *J. Chem. Phys*. **2007**, *127*, 174308.

[31] M. Lemeshko, *Phys. Rev. Lett.* **2017**, *118*, 095301.

[32] S. Grimme, *J. Comput. Chem.* **2006**, *27*, 1787-1799.

[33] A. Barducci, M. Bonomi, M. Parrinello, *WIREs Comput. Mol. Sci*. **2011**, *1*, 826-843.

[34] M. L. Strader, S. E. Feller, *J. Phys. Chem. A*. **2002**, *106*, 1074-1080.

[35] B. Aradi, B. Hourahine, Th. Frauenheim, *J. Phys. Chem. A*. **2007**, *111*, 5678-5684.

[36] S. Grimme, J. Antony, S. Ehrlich, H. Krieg, *J. Chem. Phys.* **2010**, *132*, 154104.

[37] S. Grimme, S. Ehrlich, L. Goerigk, *J. Comput. Chem.* **2011**, *32*, 1456-1465.

[38] M. J. Frisch *et al*. *Gaussian 09, rev D.01*, Gaussian, Inc., Wallingford (CT), **2009**.

[39] M. J. Abraham, T. Murtola, R. Schulz, S. Páll, J. C. Smith, B. Hess, E. Lindahl, *SoftwareX*. **2015**, *1-2*, 19-25.

[40] *ABIN, Molecular Dynamics program*. Source code available at https://github.com/photox/abin. doi:10.5281/zenodo.1228462.

[41] E. Yakaboylu, M. Lemeshko, *Phys. Rev. Lett.* **2017**, *118*, 085302.

SUPPORTING INFORMATION FOR

# Direct detection of polar structure formation in helium nanodroplets by beam deflection measurements

John W. Niman,[a] Benjamin S. Kamerin,[a] Lorenz Kranabetter,[b] Daniel J. Merthe,[a]

Jiří Suchan,[c] Petr Slavíček,[c,d] Vitaly V. Kresin[a]

[a] *Department of Physics and Astronomy, University of Southern California,
Los Angeles, CA 90089-0484, USA*

[b] *Institut für Ionenphysik und Angewandte Physik, Universität Innsbruck,
Technikerstr. 25, A-6020 Innsbruck, Austria*

[c] *Department of Physical Chemistry, University of Chemistry and Technology,
Technická 5, Prague 6, Czech Republic*

[d] *J. Heyrovský Institute of Physical Chemistry v.v.i., The Czech Academy of Sciences,
Dolejškova 3, 18223 Prague, Czech Republic*

Contents

# I. (DMSO)ₙ ion mass spectra

As described in the main text, deflection profiles of droplets doped with DMSO monomers, dimers, or trimers were acquired by setting the mass spectrometer to the masses of $(DMSO)^+$, $(DMSO)_2^+$ and $(DMSO)_3^+$ ions, respectively, and maintaining the pickup vapor pressure at a level such that the mass peak of interest would be dominant over the next higher one. This is illustrated in Fig. S1. The mass spectrometer is a Balzers QMG-511 crossed-beam quadrupole analyzer with its electron impact ionization source set to 90 eV impact energy.

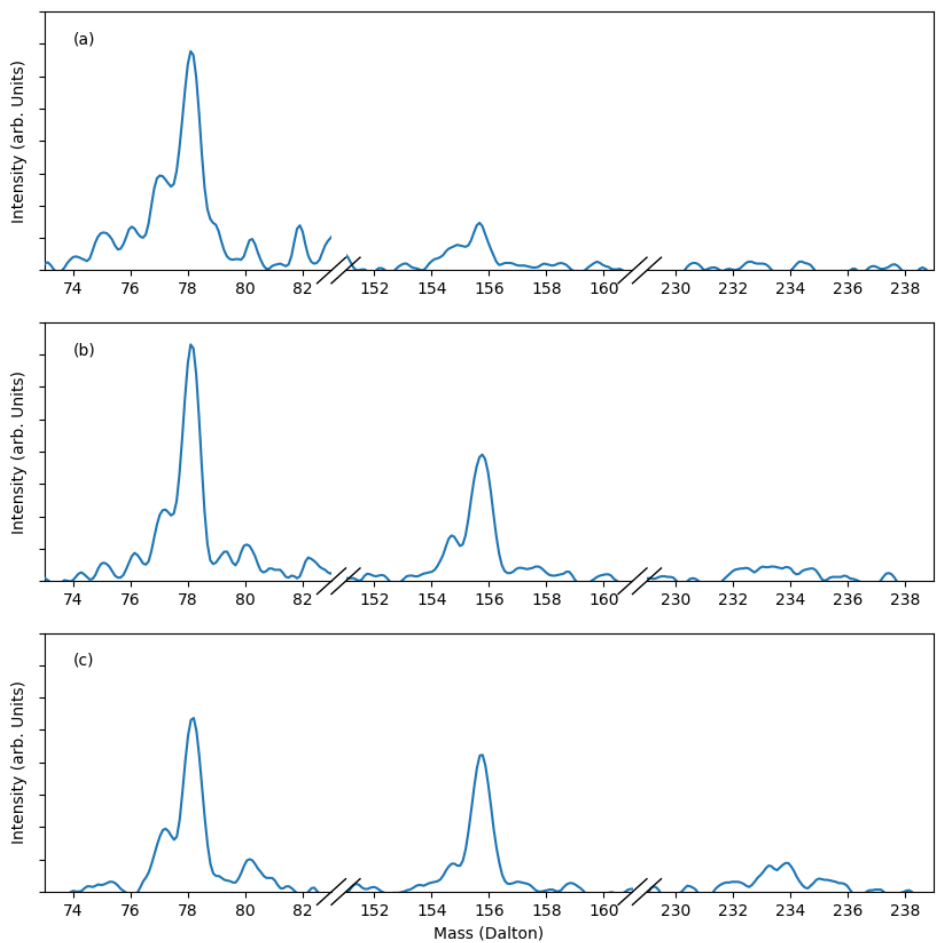

**Figure S1.** Representative mass spectra corresponding to deflection measurements on (DMSO)$_n$-doped nanodroplets. The mass spectrometer was set to the masses of intact ions: (a) 78 Da for the monomer, (b) 156 Da for the dimer, (c) 234 Da for the trimer.

## II. Ab initio calculations: Benchmarking

The potential energy surface was explored with the B3LYP(D2)/aug-cc-pVDZ method. The dipole moment of the isolated DMSO molecule in its equilibrium geometry calculated with this approach was 4.3 D, which is consistent with the tabulated value[S1] of 4.0 D within the expected accuracy of DFT.[S2] We validated this approach against the high-level CCSD(T)/aug-cc-pVTZ method. Basis set superposition error (BSSE) correction was used for all structures. The agreement is very good for all cluster structures, see Table S1. We also show the energetics of the respective minima at the DFTB/D3 level used for exploratory simulations. The DFT and CCSD(T) calculations were performed in the Gaussian 09, rev. D01 package,[S3] the DFTB results were calculated in the DFTB+ 18.2 program.[S4]

**Table S1.** Comparison of DMSO dimer binding energies at the CCSD(T), B3LYP(D2) and DFTB(D3) levels. The BSSE correction was accounted for in the CCSD(T) and B3LYP(D2) calculations.

| Dimer complex | Binding energy [eV] | | |
|---|---|---|---|
| | CCSD(T)/aug-cc-pVTZ | B3LYP/aug-cc-pVDZ D2 | DFTB |
| **D1** | 0.53 | 0.56 | 0.46 |
| **D2** | 0.46 | 0.47 | 0.41 |
| **D3** | 0.40 | 0.39 | 0.36 |
| **D4** | 0.33 | 0.33 | 0.28 |
| **D5** | 0.32 | 0.32 | 0.26 |

## III. Mapping of the (DMSO)$_2$ and (DMSO)$_3$ potential energy surfaces

The potential energy surfaces (PES) of DMSO complexes are rather rich and we mapped them in the following way. First, we performed accelerated molecular dynamics simulations with the molecular mechanics (MM) force field,[S5] using the so-called metadynamics method.[S6] Here, an additional potential is added along a preselected coordinate so that we can quickly overcome barriers along these coordinates. These simulations then also provide the free energy as a function of the selected coordinate [potential of mean force (PMF) or free energy surface (FES)]. We then selected different structures with distinct dipole moments from these metadynamical trajectories and performed further B3LYP optimization.

Metadynamics simulations were performed at 100 K to reveal the regions of interest in the dipole moment coordinate. This temperature is much higher than the experimental conditions, yet we opted for it to avoid ergodicity problems. Note that these simulations are only auxiliary, serving as a starting point for minimizations or MD simulations. The minimum on the PMF is found for a small yet non-zero dipole moment due to entropic reasons. The force field overestimates the dipole moment by 20% with respect to the *ab initio* value. The final PMFs for the dimer and trimer complexes are displayed in Fig. S2.

By clustering structures with similar dipoles together and performing 100 subsequent optimizations with Gaussian 09, for both the dimer and trimer structures, we were then able to map their PES landscapes.

The metadynamics parameters were as follows. The dimer simulation length was 100 ps, leap-frog stochastic integrator was utilized, the temperature was set to 100 K with a thermostat constant of $\tau = 1.0$ ps. For the trimer the simulation length was increased to 300 ps. The collective variable (CV) is the total dipole moment. An additional Gaussian potential was added every 100 steps. The Gaussian height was 0.015 kJ/mol and the CV gaussian width was 1.2 Debye.

MD simulations were performed with GROMACS 2018.4 code[S7] coupled with PLUMED 2.5 code[S8] for the FES simulations.

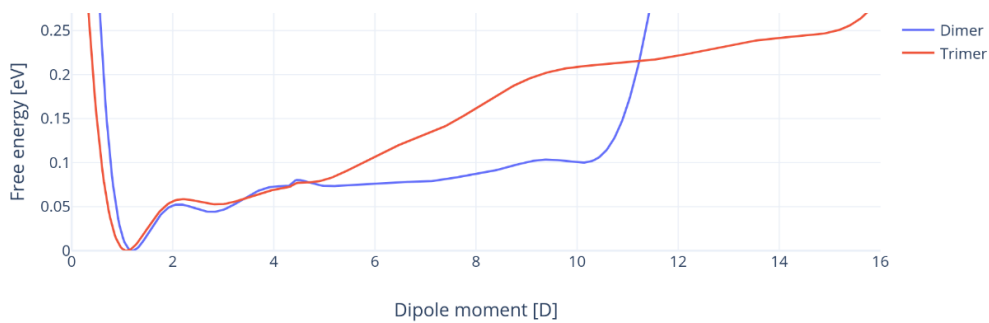

**Figure S2.** PMF for DMSO dimer and trimer complexes for the dipole moment coordinate at 100K.

## IV. Transition between two dimers at a distance and D5

Nudged elastic band (NEB) optimization[S9] was performed to find energy barriers between two DMSO molecules a distance apart (13.5 Å; in the minimal geometry at that separation the two DMSO molecules have aligned dipoles) and complex D5. Fig. S3 shows that the connection is barrierless.

The simulations were carried out in the TeraChem code[S10,S11] using the B3LYP(D2)/aug-cc-pVDZ method with 14 molecular images between the two structures. The images were generated by constrained minimization.

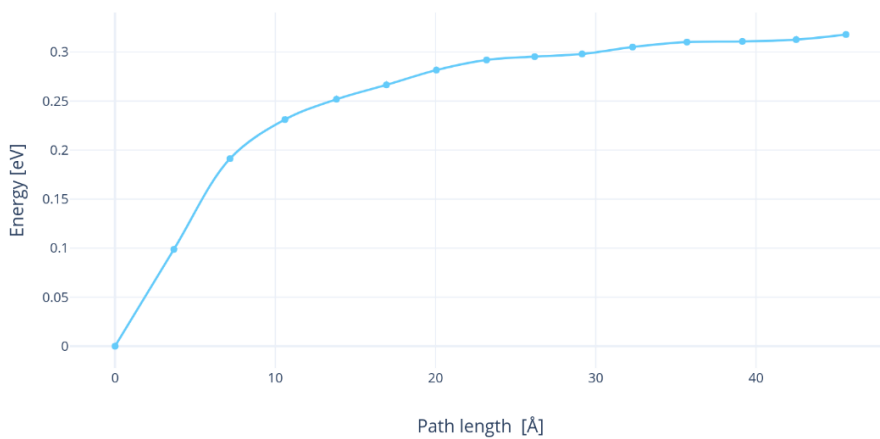

**Figure S3.** NEB calculations connecting the long-distance configuration to the D5 minimum.

## V. Transition between the D5 and D1 minima

We also performed NEB calculation connecting the D5 minimum with the global D1 minimum. The final energy curve is shown in Fig. S4.

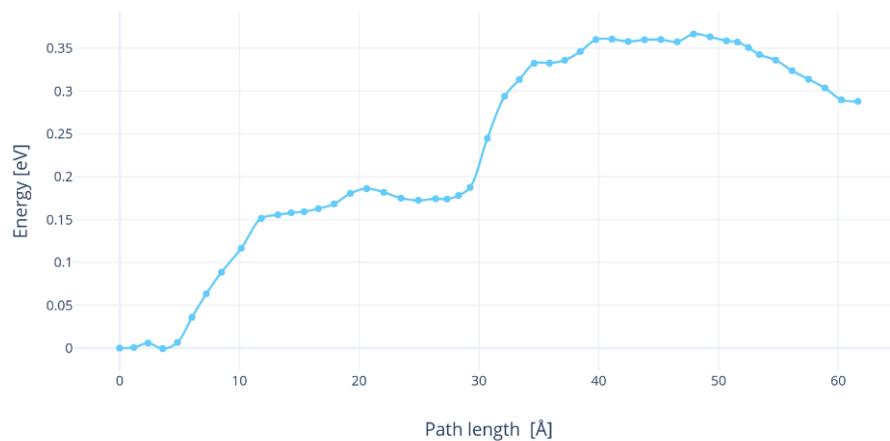

**Figure S4.** NEB calculations connecting the minima D1 and D5.

## VI. Two dimensional free energy surface

Additional insight into the topology of the multidimensional PES of DMSO aggregates can be brought about via modeling of free energy surfaces (FES). We evaluated the FES (i.e., the two dimensional version of the PMF in Fig. S2) as a function of two coordinates: the aggregate dipole moment and the interatomic S-S distance, see Figs. S5-S8. The graphs were once again generated using the metadynamics method and the temperature of 100 K to avoid convergence issues. It is clear that at large intermolecular distance the system prefers the high-dipole configuration, as mentioned above. At close distances one observes a number of minima separated by barriers.

The 2D metadynamics parameters were as follows. As before, for the dimer the simulation length was 100 ps, leap-frog stochastic integrator was utilized, the temperature was set to 100 K with thermostat constant $\tau=1.0$ ps. The first collective variable, CV1, was defined as the S-S interatomic distance between the DMSO monomers. An additional Gaussian potential was added at every 1000 steps. The Gaussian height was 0.015 kJ/mol and the CV1 Gaussian width was 0.1 nm. The second collective variable was the dipole moment with the same deposition parameters as CV1 and Gaussian width of 1.2 D. Upper energetic walls for CV1 were applied at 2 nm in order to keep the molecule in the area of interest.

For the trimer the simulation length was increased tenfold to 1000 ps, with the other parameters fixed. CV1 was redefined as the sum of S-S interatomic distances due to the presence of the third DMSO molecule, the other variables remained the same. The upper energetic walls for CV1 were shifted to 6.0 nm.

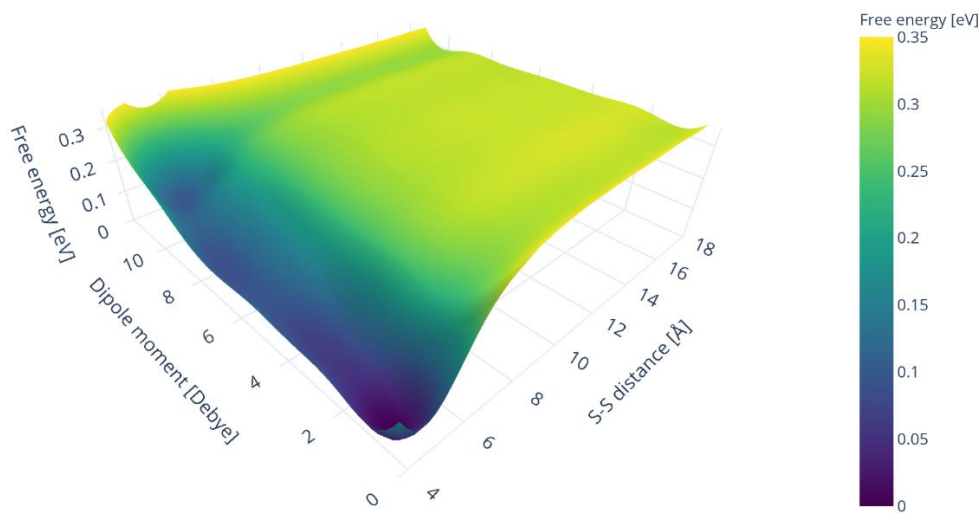

**Figure S5.** FES for the DMSO dimer at 100 K.

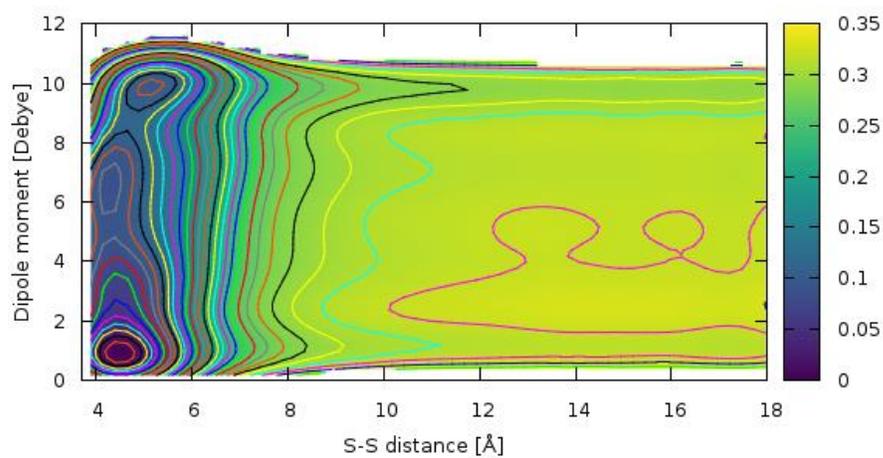

**Figure S6**. FES heatmap for the DMSO dimer at 100 K. Contour spacing 0.01 eV.

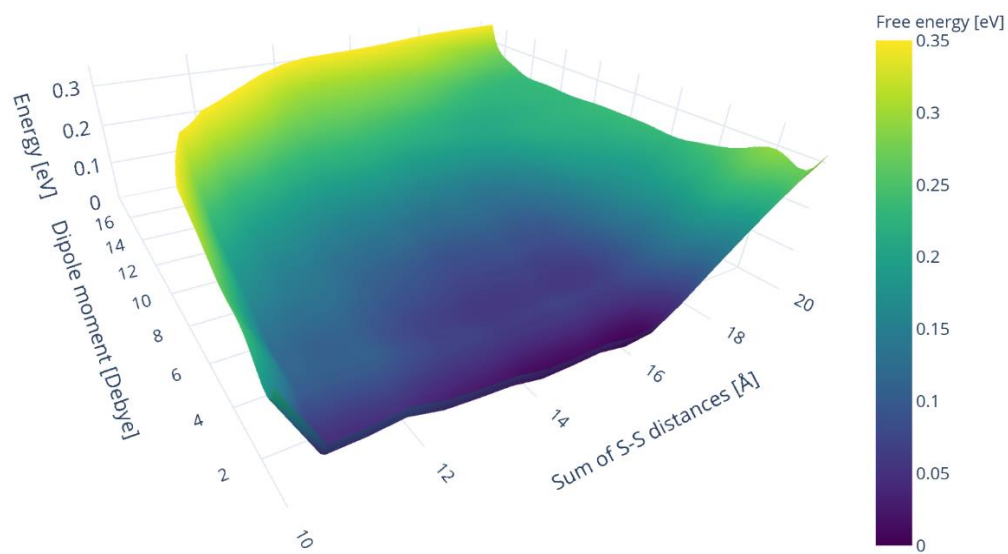

**Figure S7.** FES for the DMSO trimer at 100 K.

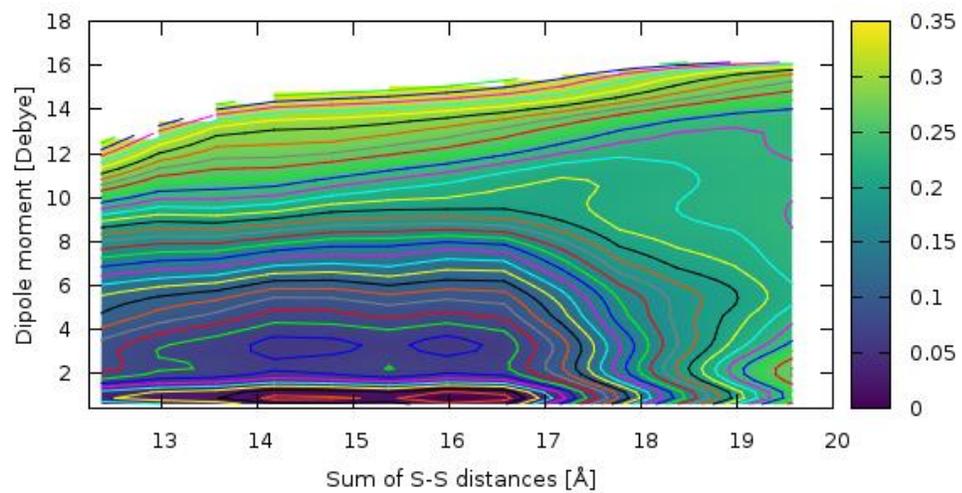

**Figure S8.** FES heatmap for the DMSO trimer at 100 K. Contour spacing 0.01 eV.

# VII. Force field parameters

The MM simulations were performed with parameters taken from ref S5. The parameters are summarized in Tables S2 and S3.

**Table S2.** Atomic type parameters for DMSO.

| Atom | Charge | $\varepsilon$ (kJ/mol) | $\sigma$ (nm) |
|:---:|:---:|:---:|:---:|
| O | -0.556 | 0.50242 | 0.30291 |
| S | 0.312 | 1.46537 | 0.35636 |
| C | -0.148 | 0.32657 | 0.36348 |
| H | 0.090 | 0.10048 | 0.23876 |

**Table S3.** Intermolecular parameters for DMSO.

| Bond | $b_0$ (nm) | $f_c$ (kJ mol$^{-1}$ nm$^{-2}$) |
|:---:|:---:|:---:|
| H-C | 0.111 | 134724.8 |
| C-S | 0.180 | 100416.0 |
| S-O | 0.153 | 225936.0 |

| Angles | $\theta_0$ (nm) | $f_c$ (kJ mol$^{-1}$ rad$^{-2}$) |
|:---:|:---:|:---:|
| H-C-H | 108.400 | 148.5320 |
| H-C-S | 111.300 | 192.8824 |
| C-S-O | 106.750 | 330.5360 |
| C-S-C | 95.000 | 142.2560 |

| Dihedrals | $\varphi_0$ (deg) | $f_c$ (kJ mol$^{-1}$) | X |
|:---:|:---:|:---:|:---:|
| H-C-S-O | 0.0 | 0.8368 | 3 |
| H-C-S-C | 0.0 | 0.8368 | 3 |

# VIII. Cartesian coordinates of all structures

Geometries of the optimal structures presented in Figs. 2 and 4 of the main text are listed below, with all coordinates in Angstroms.

Monomer
 10

| | | | |
|---|---|---|---|
| C | 1.390750 | 0.279323 | -0.278296 |
| S | 0.072728 | -0.679506 | 0.585004 |
| C | -1.342363 | 0.171624 | -0.236526 |
| O | 0.075720 | -0.189227 | 2.044957 |
| H | 1.346127 | 0.069130 | -1.356470 |
| H | 1.227420 | 1.344812 | -0.066434 |
| H | 2.347951 | -0.053762 | 0.141464 |
| H | -1.314193 | -0.035613 | -1.315827 |
| H | -2.257159 | -0.235280 | 0.211752 |
| H | -1.256995 | 1.246897 | -0.028367 |

D1
 20

| | | | |
|---|---|---|---|
| C | 1.391830 | 0.296071 | -0.267191 |
| S | 0.073035 | -0.689518 | 0.555023 |
| C | -1.343411 | 0.188626 | -0.225497 |
| O | 0.078854 | -0.256270 | 2.047129 |
| H | 1.379923 | 0.056354 | -1.339642 |
| H | 1.179380 | 1.359143 | -0.087564 |
| H | 2.341205 | -0.018939 | 0.183566 |
| H | -1.345350 | -0.050790 | -1.298081 |
| H | -2.251051 | -0.199163 | 0.253587 |
| H | -1.209576 | 1.265306 | -0.051326 |
| S | -0.073081 | 3.587139 | 1.938615 |
| O | -0.078682 | 3.153903 | 0.446506 |
| C | -1.391839 | 2.601376 | 2.760682 |
| C | 1.343385 | 2.709142 | 2.719264 |
| H | -1.380080 | 2.841096 | 3.833135 |
| H | -2.341208 | 2.916257 | 2.309824 |
| H | -1.179218 | 1.538334 | 2.581077 |
| H | 1.345162 | 2.948483 | 3.791865 |
| H | 1.209699 | 1.632459 | 2.545001 |
| H | 2.251033 | 3.097074 | 2.240313 |

D2
 20

| | | | |
|---|---|---|---|
| O | 8.826895 | 8.110270 | 10.400227 |
| S | 9.972464 | 7.608129 | 9.483674 |
| C | 11.539269 | 7.974326 | 10.379345 |
| H | 12.382316 | 7.730081 | 9.717863 |
| H | 11.525836 | 9.039717 | 10.646509 |
| H | 11.551024 | 7.328890 | 11.266521 |
| C | 10.186293 | 8.892506 | 8.183524 |
| H | 11.035534 | 8.602394 | 7.548535 |
| H | 9.256912 | 8.900193 | 7.600878 |
| H | 10.362717 | 9.852789 | 8.687985 |
| O | 10.697510 | 11.161246 | 10.536141 |
| S | 9.253455 | 11.346438 | 11.076378 |
| C | 9.145789 | 10.330797 | 12.600760 |
| H | 8.213528 | 10.584098 | 13.124409 |
| H | 9.133390 | 9.289859 | 12.256910 |
| H | 10.027135 | 10.559985 | 13.215271 |
| C | 9.274676 | 12.983721 | 11.916622 |
| H | 8.313117 | 13.137349 | 12.426572 |
| H | 10.112562 | 12.992963 | 12.626663 |
| H | 9.422581 | 13.740250 | 11.136213 |

D3
 20

| | | | |
|---|---|---|---|
| C | -6.599507 | -9.935427 | -8.808258 |
| S | -6.050826 | -8.220930 | -8.418528 |
| C | -7.587463 | -7.381111 | -8.969174 |
| O | -4.975044 | -7.895873 | -9.486073 |
| H | -7.454125 | -10.195223 | -8.167596 |
| H | -6.867985 | -9.974321 | -9.872589 |
| H | -5.748550 | -10.596439 | -8.602335 |
| H | -8.424339 | -7.741197 | -8.354184 |
| H | -7.407231 | -6.311806 | -8.806702 |
| H | -7.734957 | -7.620273 | -10.031201 |
| S | -4.736135 | -4.653480 | -8.965041 |
| O | -5.810347 | -4.978975 | -7.896071 |
| C | -4.188088 | -2.938502 | -8.576590 |
| C | -3.198394 | -5.492320 | -8.416137 |
| H | -3.333958 | -2.678643 | -9.217880 |
| H | -5.039455 | -2.277959 | -8.782326 |
| H | -3.918996 | -2.898989 | -7.512438 |
| H | -2.362832 | -5.133640 | -9.033730 |
| H | -3.048402 | -5.251363 | -7.354870 |
| H | -3.379470 | -6.561814 | -8.576431 |

D4
 20

O   10.770543   12.406453   7.172925
S   10.996603   12.217603   8.692035
C   12.663048   11.452412   8.865018
H   12.827871   11.198932   9.921240
H   12.692858   10.559951   8.226519
H   13.389878   12.202747   8.531247
C   10.042817   10.713973   9.163143
H   10.255542   10.475201   10.214216
H    8.982319   10.960304   9.031380
H   10.346341    9.897937   8.494552
O   11.783311    8.009573   5.181466
S   11.481115    9.357691   5.867285
C   12.445813   10.642888   4.970928
H   12.145669   11.622494   5.364081
H   13.505277   10.435700   5.167309
H   12.228607   10.534564   3.900118
C    9.828399    9.903572   5.270228
H    9.653478   10.918986   5.647899
H    9.840437    9.858877   4.173228
H    9.097072    9.191144   5.672183

D5
 20

C    0.832083    0.053205   0.687160
S   -0.018838    0.529481  -0.873482
S    1.568387   -2.573673  -1.989863
C    0.721902   -2.067199  -3.541948
C   -1.624708   -0.272902  -0.470355
O   -0.252453    2.053432  -0.806970
O    0.449839   -2.770887  -0.938584
C    2.024752   -4.266109  -2.547539
H    0.867295   -1.043176   0.726962
H    1.838413    0.487869   0.644560
H    0.262499    0.486055   1.520395
H   -1.455444   -1.355871  -0.418903
H   -1.978107    0.143628   0.482340
H   -2.317341   -0.012725  -1.280529
H    2.733390   -4.189503  -3.384114
H    2.491453   -4.765940  -1.690035
H    1.103863   -4.786018  -2.843846
H    1.450590   -2.071775  -4.364292
H   -0.098565   -2.774219  -3.724319
H    0.340792   -1.054444  -3.366810

T1
 30

O  18.325573  19.811653  21.617481
S  19.721884  20.312996  21.169378
C  20.228323  21.608815  22.376809
H  21.215482  21.985848  22.075579
H  19.485068  22.415686  22.369997
H  20.289383  21.113897  23.353709
C  19.440363  21.463757  19.758658
H  20.418372  21.838813  19.426676
H  18.968548  20.870817  18.965678
H  18.799604  22.289633  20.091628
O  18.271096  24.146530  21.392567
S  17.718310  25.601134  21.478337
C  16.938575  25.747636  23.136233
H  16.505635  26.754394  23.216988
H  16.166283  24.969074  23.217754
H  17.742714  25.618895  23.871189
C  16.151632  25.602873  20.517245
H  15.714781  26.608754  20.587075
H  16.422965  25.376776  19.478677
H  15.486136  24.843390  20.952583
O  14.642326  23.527870  22.517495
S  15.116812  22.043647  22.457649
C  16.675088  21.971241  23.424806
H  17.142014  20.993610  23.248081
H  17.321498  22.778503  23.058847
H  16.399735  22.113636  24.476868
C  15.887654  21.824953  20.806281
H  16.405169  20.856910  20.796220
H  15.074819  21.866766  20.071118
H  16.603161  22.645493  20.671394

T2
 30

C  19.978371  22.960022  19.710536
S  19.735892  23.242403  21.510328
C  20.388685  21.613064  22.058900
O  18.198954  23.189838  21.745112
C  15.358815  23.917374  20.096399
S  14.121084  23.111607  21.194225
C  15.047600  23.377434  22.762579
O  14.190395  21.590951  20.897064
S  17.054341  20.556144  20.208772
C  15.942352  19.217124  19.628149

| | | | |
|---|---|---|---|
| O | 18.501515 | 20.064915 | 19.893824 |
| C | 16.827935 | 20.251673 | 22.004302 |
| H | 21.059208 | 22.936455 | 19.512556 |
| H | 19.496985 | 22.004570 | 19.454354 |
| H | 19.514939 | 23.811524 | 19.197357 |
| H | 21.467010 | 21.591478 | 21.848409 |
| H | 20.210845 | 21.552465 | 23.139753 |
| H | 19.850741 | 20.831749 | 21.503192 |
| H | 15.760191 | 20.345461 | 22.229074 |
| H | 17.216281 | 19.244890 | 22.206490 |
| H | 17.413863 | 21.021960 | 22.516917 |
| H | 14.919373 | 19.492859 | 19.911972 |
| H | 16.052332 | 19.168894 | 18.537795 |
| H | 16.267612 | 18.274891 | 20.088960 |
| H | 15.332881 | 24.997756 | 20.296283 |
| H | 16.349231 | 23.496971 | 20.310933 |
| H | 15.040118 | 23.707847 | 19.068112 |
| H | 15.045690 | 24.454363 | 22.980238 |
| H | 14.504377 | 22.828117 | 23.541197 |
| H | 16.073387 | 23.010200 | 22.639216 |

T3
 30

| | | | |
|---|---|---|---|
| C | 19.318080 | 22.762136 | 22.418743 |
| S | 18.401131 | 21.379365 | 21.618767 |
| C | 18.852818 | 21.822207 | 19.888771 |
| O | 16.894078 | 21.701134 | 21.778298 |
| O | 19.025390 | 25.162225 | 20.116170 |
| S | 18.978878 | 26.478380 | 19.285963 |
| C | 18.218864 | 27.737277 | 20.375610 |
| C | 20.695828 | 27.130948 | 19.340978 |
| C | 16.290257 | 24.915699 | 22.126303 |
| S | 15.070039 | 25.136098 | 20.770491 |
| C | 15.850747 | 23.953991 | 19.603056 |
| O | 15.323351 | 26.555016 | 20.184537 |
| H | 20.393691 | 22.575593 | 22.293887 |
| H | 19.028486 | 23.706104 | 21.941314 |
| H | 19.045684 | 22.740531 | 23.480909 |
| H | 19.921248 | 21.605360 | 19.751253 |
| H | 18.245550 | 21.181412 | 19.238001 |
| H | 18.646287 | 22.887366 | 19.725303 |
| H | 20.713003 | 28.127033 | 18.877545 |
| H | 21.008893 | 27.173273 | 20.392785 |
| H | 21.324967 | 26.432136 | 18.776249 |
| H | 18.326517 | 28.718024 | 19.891996 |
| H | 17.157993 | 27.455148 | 20.467183 |
| H | 18.745592 | 27.706182 | 21.339379 |
| H | 16.277900 | 23.860237 | 22.428986 |
| H | 17.274927 | 25.192098 | 21.728825 |
| H | 15.980785 | 25.581780 | 22.940796 |
| H | 15.862931 | 22.963479 | 20.077728 |
| H | 15.245255 | 23.967205 | 18.688986 |
| H | 16.872837 | 24.307564 | 19.416185 |

T4
 30

| | | | |
|---|---|---|---|
| C | 5.071517 | -0.034670 | -3.655985 |
| S | 4.287297 | -0.194668 | -1.997307 |
| C | 5.830416 | -0.559251 | -1.075361 |
| O | 3.893217 | 1.257561 | -1.611793 |
| S | -0.020529 | 2.540318 | 0.295084 |
| C | 1.283159 | 3.066414 | -0.885371 |
| O | 0.572434 | 2.732177 | 1.715202 |
| C | 3.644181 | 1.756288 | 1.598188 |
| S | 2.617946 | 0.269497 | 1.276626 |
| C | 1.964214 | 0.086045 | 2.982365 |
| O | 3.615924 | -0.905124 | 1.078908 |
| H | 1.323627 | 0.953582 | 3.183548 |
| H | 1.388114 | -0.847471 | 2.998202 |
| H | 2.818493 | 0.022048 | 3.669544 |
| H | 2.968760 | 2.565559 | 1.900207 |
| H | 4.362207 | 1.495270 | 2.386815 |
| H | 4.148555 | 1.979398 | 0.651252 |
| H | 0.822989 | 3.182558 | -1.876460 |
| H | 1.688760 | 4.020594 | -0.522132 |
| H | 2.057413 | 2.289562 | -0.920766 |
| H | 5.497260 | -1.005416 | -3.946611 |
| H | 4.282043 | 0.264332 | -4.356496 |
| H | 5.846821 | 0.740863 | -3.594728 |
| H | 6.247662 | -1.502562 | -1.454190 |
| H | 6.520679 | 0.279922 | -1.237121 |
| H | 5.523931 | -0.655551 | -0.027295 |
| C | -1.126902 | 3.988674 | 0.041019 |
| H | -1.509237 | 3.972861 | -0.989222 |
| H | -1.948976 | 3.889460 | 0.760368 |
| H | -0.546405 | 4.899313 | 0.241051 |

T5
 30

O   16.634857   24.592189   18.933949
S   17.819686   24.188608   19.842340
C   18.546611   25.748060   20.487096
H   19.438384   25.480875   21.069767
H   18.776646   26.394312   19.630220
H   17.780785   26.213377   21.119207
C   19.237765   23.746631   18.763679
H   20.096423   23.565749   19.425082
H   18.955755   22.836787   18.219781
H   19.417204   24.574596   18.065138
O   20.209226   23.485765   22.163446
S   18.877554   23.104769   22.864853
C   19.254372   22.812446   24.636229
H   18.324234   22.468296   25.107880
H   20.058270   22.067975   24.708308
H   19.577739   23.772502   25.056335
C   18.574534   21.347602   22.430705
H   17.751416   20.981712   23.056223
H   18.299881   21.331231   21.369829
H   19.512448   20.803987   22.601658
O   15.856035   22.403783   23.805490
S   15.408064   22.687282   22.347855
C   15.384240   24.514743   22.152187
H   15.163099   24.748814   21.101020
H   14.647435   24.936357   22.847504
H   16.395863   24.844707   22.412702
C   13.585634   22.453003   22.322482
H   13.200021   22.813677   21.358789
H   13.395364   21.379488   22.440968
H   13.163459   23.017678   23.164279

T6
 30

O   18.248791   19.538962   25.445555
S   18.177691   19.442715   26.988687
C   18.433034   21.156221   27.613792
H   18.496857   21.127477   28.710183
H   19.356533   21.546423   27.166533
H   17.561677   21.740741   27.294830
C   19.813685   18.792657   27.529990
H   19.853766   18.804080   28.627823
H   19.884199   17.765020   27.153619
H   20.593531   19.428978   27.091708

| | | | |
|---|---|---|---|
| O | 22.122928 | 22.013981 | 19.259183 |
| S | 22.029013 | 21.907157 | 20.795300 |
| C | 22.263129 | 23.603652 | 21.467768 |
| H | 22.337954 | 23.520943 | 22.560161 |
| H | 23.173825 | 24.017982 | 21.015392 |
| H | 21.385740 | 24.187672 | 21.162752 |
| C | 23.639059 | 21.242274 | 21.387621 |
| H | 23.636418 | 21.292751 | 22.484521 |
| H | 23.704654 | 20.207811 | 21.027740 |
| H | 24.435317 | 21.852934 | 20.941810 |
| O | 22.288179 | 21.938705 | 24.307074 |
| S | 21.002813 | 21.173364 | 24.716842 |
| C | 20.992990 | 19.611911 | 23.748124 |
| H | 20.049272 | 19.097423 | 23.964572 |
| H | 21.091252 | 19.874277 | 22.687275 |
| H | 21.857084 | 19.027672 | 24.087681 |
| C | 19.609864 | 21.978664 | 23.829299 |
| H | 18.704772 | 21.397957 | 24.043202 |
| H | 19.535878 | 22.999435 | 24.224070 |
| H | 19.855097 | 21.990158 | 22.759892 |

T7
 30

| | | | |
|---|---|---|---|
| C | 0.472180 | -0.265895 | 0.733682 |
| S | -0.136458 | 0.321670 | -0.898489 |
| O | -0.384447 | 1.846400 | -0.754727 |
| C | -1.785515 | -0.482541 | -0.798695 |
| O | 0.359320 | -2.893610 | -1.176365 |
| S | 1.641205 | -2.693731 | -2.021010 |
| C | 2.115524 | -4.372107 | -2.603230 |
| C | 1.077057 | -2.055772 | -3.650613 |
| H | 0.551026 | -1.358283 | 0.675742 |
| H | 1.451728 | 0.200982 | 0.892637 |
| H | -0.247228 | 0.058360 | 1.495983 |
| H | -1.625451 | -1.566818 | -0.828784 |
| H | -2.261186 | -0.155086 | 0.134675 |
| H | -2.355211 | -0.137931 | -1.670337 |
| H | 2.951015 | -4.283369 | -3.311612 |
| H | 2.420451 | -4.939134 | -1.715311 |
| H | 1.234961 | -4.831777 | -3.071533 |
| H | 1.927235 | -2.048859 | -4.346467 |
| H | 0.268384 | -2.708093 | -4.006330 |
| H | 0.713231 | -1.036965 | -3.472756 |
| S | -2.320128 | 1.993864 | 2.169713 |
| O | -3.154515 | 2.294445 | 3.432453 |
| C | -3.077491 | 2.942538 | 0.787412 |
| C | -0.788980 | 3.007148 | 2.285051 |
| H | -2.419933 | 2.841894 | -0.086210 |
| H | -4.064898 | 2.499581 | 0.606587 |
| H | -3.180753 | 3.985061 | 1.116591 |
| H | -0.254575 | 2.912504 | 1.330540 |
| H | -1.090007 | 4.041603 | 2.496975 |
| H | -0.204794 | 2.601451 | 3.120616 |

T8

| | | | |
|---|---|---|---|
| O | 22.134918 | 25.903523 | 23.940447 |
| S | 23.041269 | 24.808338 | 24.561304 |
| C | 22.070844 | 24.071469 | 25.935851 |
| H | 22.726180 | 23.402398 | 26.509576 |
| H | 21.686789 | 24.890733 | 26.558087 |
| H | 21.251172 | 23.507444 | 25.474200 |
| C | 24.246540 | 25.712157 | 25.614598 |
| H | 24.837479 | 24.984221 | 26.187314 |
| H | 24.891884 | 26.286350 | 24.938904 |
| H | 23.682050 | 26.382530 | 26.276309 |
| O | 18.026224 | 22.657810 | 23.136880 |
| S | 19.327282 | 23.413547 | 23.474273 |
| C | 18.847931 | 25.002640 | 24.270183 |
| H | 19.769957 | 25.568621 | 24.458552 |
| H | 18.190563 | 25.554032 | 23.588146 |
| H | 18.338681 | 24.736761 | 25.204821 |
| C | 19.960394 | 24.149959 | 21.912549 |
| H | 20.865655 | 24.718669 | 22.164650 |
| H | 20.186067 | 23.308764 | 21.246021 |
| H | 19.190316 | 24.810736 | 21.496604 |
| O | 18.606089 | 27.152941 | 21.810753 |
| S | 19.594310 | 28.340832 | 21.800217 |
| C | 21.206535 | 27.678398 | 21.199666 |
| H | 21.938315 | 28.498325 | 21.206754 |
| H | 21.524346 | 26.861633 | 21.859709 |
| H | 21.033168 | 27.327994 | 20.175000 |
| C | 20.127602 | 28.590982 | 23.547016 |
| H | 20.862715 | 29.407639 | 23.569519 |
| H | 19.226795 | 28.869318 | 24.107320 |
| H | 20.565929 | 27.657234 | 23.920780 |

References

[S1] *CRC Handbook of Chemistry and Physics*, 99th ed. (Ed.: J. R. Rumble), CRC Press, Boca Raton, **2018**.

[S2] D. Hait, M. Head-Gordon, *J. Chem. Theory Comput.* **2018**, *14*, 1969.

[S3] M. J. Frisch, G. W. Trucks, H. B. Schlegel, G. E.Scuseria, M. A. Robb, J. R. Cheeseman, G. Scalmani, V. Barone, B. Mennucci, G. A. Petersson, H. Nakatsuji, M. Caricato, X. Li, H. P. Hratchian, A. F. Izmaylov, J. Bloino, G. Zheng, J. L. Sonnenberg, M. Hada, M. Ehara, K. Toyota, R. Fukuda, J. Hasegawa, M. Ishida, T. Nakajima, Y. Honda, O. Kitao, H. Nakai, T. Vreven, J. A. Montgomery, Jr., J. E. Peralta, F. Ogliaro, M. Bearpark, J. J. Heyd, E. Brothers, K. N. Kudin, V. N. Staroverov, R. Kobayashi, J. Normand, K. Raghavachari, A. Rendell, J. C. Burant, S. S. Iyengar, J. Tomasi, M. Cossi, N. Rega, J. M. Millam, M. Klene, J. E. Knox, J. B. Cross, V. Bakken, C. Adamo, J. Jaramillo, R. Gomperts, R. E. Stratmann, O. Yazyev, A. J. Austin, R. Cammi, C. Pomelli, J. W. Ochterski, R. L. Martin, K. Morokuma, V. G. Zakrzewski, G. A. Voth, P. Salvador, J. J. Dannenberg, S. Dapprich, A. D. Daniels, Ö. Farkas, J. B. Foresman, J. V. Ortiz, J. Cioslowski, D. J. Fox, *Gaussian 09, rev. D.01*, Gaussian, Inc., Wallingford (CT), **2009**.

[S4] B. Aradi, B. Hourahine, Th. Frauenheim, *J. Phys. Chem. A*. **2007**, *111*, 5678.

[S5] M. L.Strader, S. E. Feller, *J. Phys. Chem. A*. **2002**, *106*, 1074.

[S6] A. Barducci, M. Bonomi, M. Parrinello, *WIREs Comput. Mol. Sci*. **2011**, *1*, 826.

[S7] M. J. Abraham, T. Murtola, R. Schulz, S. Páll, J. C. Smith, B. Hess, E. Lindahl, *SoftwareX* **2015**, *1-2*, 19.

[S8] G. A. Tribello, M. Bonomi, D. Branduardi, C. Camilloni, G. Bussi, *Comput. Phys. Commun*. **2014**, *185*, 604.

[S9] H. Jonson, G. Mills, K. W. Jacobsen, in *Classical and Quantum Dynamics in Condensed Phase Simulations* (Eds.: B. J. Berne, G. Ciccotti, D. F. Coker, D. F.), World Scientific, Singapore, **1998**, pp. 385-404.

[S10] TeraChem v 1.9. http://www.petachem.com.

[S11] I. S. Ufimtsev, T. J. Martinez, *J. Chem. Theory Comput.* **2009**, *5*, 2619.